# Bi-confluent Heun potentials for a stationary relativistic wave equation for a spinless particle

**H.H. Azizbekyan[1,2], A.M. Manukyan[1,2], V.M. Mekhitarian[1,2], and A.M. Ishkhanyan[1,2,3]**

[1]Russian-Armenian University, Yerevan 0051, Armenia
[2]Institute for Physical Research, Ashtarak 0203, Armenia
[3]Armenian State Pedagogical University, Yerevan 0010, Armenia

The variety of bi-confluent Heun potentials for a stationary relativistic wave equation for a spinless particle is presented. The physical potentials and energy spectrum of this wave equation are related to those for a corresponding Schrödinger equation in the sense that all the potentials derived for the latter equation are also applicable for the wave equation under consideration. We show that in contrast to the Schrödinger equation the characteristic spatial length of the potential imposes a restriction on the energy spectrum that directly reflects the uncertainty principle. Studying the inverse-square-root bi-confluent Heun potential, it is shown that the uncertainty principle limits, from below, the principal quantum number for the bound states, i.e., physically feasible states have an *infimum* cut so that the ground state adopts a higher quantum number as compared to the Schrödinger case.

## 1. Introduction

We consider the bi-confluent Heun potentials for a stationary relativistic wave equation for a spinless particle (RWE) [1]. This equation, which is obtained by applying the invariance principle for the total four-dimensional momentum of the system "field+particle", has some significant advantages as compared to its analog equations such as the Klein-Gordon and Dirac equations [2,3]. For instance, for the problem of a hydrogen-like atom, it has solutions for an arbitrary value of the interaction constant not restricted to whatever atomic number of the nucleus (we recall that for the Dirac equation the atomic number is restricted to $Z < 137$ [4]). In contrast to the Schrödinger equation, the energy spectrum of the ground state for the considered equation proves to be limited by a spatial characteristic size. This property directly reflects the uncertainty principle in that, irrespective of the well depth value, the particle can be localized in a bound state only if the well width is larger than the half-wavelength of the particle. The equation is applicable for different types of particles and interactions. The analysis of the solutions shows the full compliance with the principles of relativistic and quantum mechanics, and the solutions are devoid of any restrictions on the nature and magnitude of the interactions.

In the present paper, we examine the properties of the solutions of the mentioned wave equation by reducing it to the bi-confluent Heun equation [5] via duly chosen transformation of the independent and dependent variables [6]. This leads to exact solutions in terms of the bi-confluent Heun functions for five independent six-parameter potentials [7,8]. Among these, there exists a particular exactly solvable sub-potential for which the solution is written in terms of two Hermite functions which are much more studied functions than the bi-confluent Heun functions. This is the inverse-square-root potential which is of a long-range type to sustain an infinite number of bound states [9]. Discussing the properties of the spectrum for this potential, we show that the spectrum for relativistic wave equation under consideration reveals an *infimum* cut, i.e., the ground state generally adopts a higher quantum number as compared with the Schrödinger spectrum.

## 2. The relativistic wave equation and the Schrödinger equation

We consider the following one-dimensional stationary relativistic wave equation (RWE) for a spinless particle of mass $m$ and energy $E$ in the potential field $V(x)$ [1]:

$$\frac{d^2\psi}{dx^2} + \frac{W^2 - \left(mc^2 + q_0\varphi(x)\right)^2 + q_0^2\mathbf{A}^2(x)}{\hbar^2 c^2}\psi = 0, \quad (1)$$

where $\hbar$ is the reduced Planck constant, $c$ is the speed of light, $W$ is the energy of the particle, $q_0$ is the charge of the particle, and the functions $\varphi$ and $\mathbf{A}$ are the scalar and vector potentials, respectively. With notations

$$\frac{W^2 - m^2c^4}{2mc^2} = E, \quad (2)$$

and
$$q_0\varphi(x) + \frac{q_0^2}{2mc^2}\left(\varphi^2(x) - \mathbf{A}^2(x)\right) = V(x) \quad (3)$$

this equation is conveniently rewritten in the form of the familiar Schrödinger equation:

$$\frac{d^2\psi}{dx^2} + + \frac{2m}{\hbar^2}\left(E - V(x)\right)\psi = 0. \quad (4)$$

Equation (3) indicates that diverse combinations of scalar and vector potentials $\varphi$ and $\mathbf{A}$ may correspond to the same Schrödinger potential $V(x)$. In the case of consideration of the vector potential alone, it obeys the relation

$$q_0^2\mathbf{A}^2(x) = -2mc^2 V(x), \quad V(x) < 0. \quad (5)$$

Resolving equation (3) with respect to $\varphi$, the scalar potential is presented as

$$q_0\varphi(x) = \left(V(x) + \frac{q_0^2}{2mc^2}\mathbf{A}^2(x)\right)\frac{2}{1\pm\sqrt{1+\frac{2}{mc^2}\left(V(x)+\frac{q_0^2}{2mc^2}\mathbf{A}^2(x)\right)}}. \qquad (6)$$

Hence, if the vector potential is not included for a particular problem then the scalar potential is given by the Schrödinger potential as

$$q_0\varphi(x) = V(x)\frac{2}{1\pm\sqrt{1+\frac{2V(x)}{mc^2}}}, \quad V(x) > -\frac{mc^2}{2}. \qquad (7)$$

## 3. Bi-confluent Heun potentials

There are five six-parameter potentials for which the general solution of the one-dimensional stationary Schrödinger equation is written in terms of the bi-confluent Heun functions [6-8]. These potentials are obtained as follows. Applying the change of the independent variable $z = z(x)$ followed by the transformation of the dependent variable $\psi = \theta(z)u(z)$, the Schrödinger equation is reduced to an equation for the new dependent variable $u(z)$ written as

$$\frac{d^2u}{dz^2} + \left(2\frac{\theta_z}{\theta}+\frac{\rho_z}{\rho}\right)\frac{du}{dz} + \left(\frac{\theta_{zz}}{\theta}+\frac{\rho_z}{\rho}\frac{\theta_z}{\theta}+\frac{2m}{\hbar^2}\frac{E-V(z)}{\rho^2}\right)u = 0, \qquad (8)$$

where $\rho = dz/dx$. Demanding this equation to become the bi-confluent Heun equation [5]:

$$\frac{d^2u}{dz^2} + \left(\frac{\gamma}{z}+\delta+\varepsilon z\right)\frac{du}{dz} + \frac{\alpha z - q}{z}u = 0, \qquad (9)$$

one arrives at the following system of two coupled equations:

$$2\frac{\theta_z}{\theta}+\frac{\rho_z}{\rho} = \frac{\gamma}{z}+\delta+\varepsilon z \qquad (10)$$

and

$$\frac{\theta_{zz}}{\theta}+\frac{\rho_z}{\rho}\frac{\theta_z}{\theta}+\frac{2m}{\hbar^2}\frac{E-V(z)}{\rho^2} = \frac{\alpha z - q}{z}. \qquad (11)$$

According to arguments of [8], the solution of these equations is constructed by putting $\rho = z^{m_1}$ with an integer or half-integer $m_1$. Then, resolving equation (10) with respect to $\theta$ and substituting $\theta$ into equation (11), one arrives at five independent bi-confluent Heun potentials first presented by Lemieux and Bose [7]. For reader's convenience, these potentials are represented here in Table 1. We not that all five potentials are six-parameter because $V_{0,1,2,3,4}$ are arbitrary and one may replace the coordinate $x$ by $x-x_0$ with arbitrary $x_0$.

Furthermore, all potential parameters may be chosen complex. This is a useful point because one may then construct $PT$-symmetric non-Hermitian potentials [10,11]. In the present paper, however, we assume the parameters $V_{0,1,2,3,4}$ real and put $x_0 = 0$ for convenience.

| $m_1$ | Bi-confluent Heun potential $V(x)$ | Coordinate transformation |
|---|---|---|
| $-1$ | $V_0 + \dfrac{V_1}{\sqrt{x}} + \dfrac{V_2}{x} + \dfrac{V_3}{x^{3/2}} + \dfrac{V_4}{x^2}$ | $z = \sqrt{2x}$ |
| $-1/2$ | $V_0 + V_1 x^{2/3} + \dfrac{V_2}{x^{2/3}} + \dfrac{V_3}{x^{4/3}} + \dfrac{V_4}{x^2}$ | $z = (3x/2)^{2/3}$ |
| $0$ | $V_0 + V_1 x + V_2 x^2 + \dfrac{V_3}{x} + \dfrac{V_4}{x^2}$ | $z = x$ |
| $1/2$ | $V_0 + V_1 x^2 + V_2 x^4 + V_3 x^6 + \dfrac{V_4}{x^2}$ | $z = x^2/4$ |
| $1$ | $V_0 + V_1 e^x + V_2 e^{2x} + V_3 e^{3x} + V_4 e^{4x}$ | $z = e^x$ |

Table 1. Five six-parameter bi-confluent Heun potentials ($V_{0,1,2,3,4}$ are arbitrary constants) together with the corresponding coordinate transformation $z = z(x)$ [6-8].

The solution of the stationary Schrödinger equation (4) for the bi-confluent Heun potentials is explicitly written in terms of the bi-confluent Heun function $H_B$ as

$$\psi = z^{\alpha_0} e^{\alpha_1 z + \alpha_2 z^2} H_B(\gamma, \delta, \varepsilon; \alpha, q; z). \qquad (12)$$

Starting from a particular potential of Table 1, one calculates, through equations presented in [8], the parameters $\gamma, \delta, \varepsilon, \alpha, q$ of $H_B$ and the parameters $\alpha_{0,1,2}$ of the pre-factor of solution (12). It should be noted, however, that the bi-confluent Heun function is a rather complicated mathematical object to handle [12,13]. To overcome this, one may apply a quite recently developed technique of expansion of the bi-confluent Heun function in terms of the familiar Hermite functions [8]. In this way, it is possible to identify a particular exactly solvable sub potential for which the expansion involves just two noninteger-order Hermite functions of a scaled and shifted argument. This is the inverse-square-root potential [9]

$$V = \dfrac{V_0}{\sqrt{x}} \qquad (13)$$

with arbitrary (real or complex) interaction strength $V_0$.

## 4. Energy spectrum for the inverse-square-root potential

The inverse-square-root potential is a long-range potential so that its attractive version (realized if $V_0 < 0$) sustains an infinite number of bound states [9]. To discuss the energy spectrum for these states, we note that, according to equation (2), the spectrum found for the Schrödinger equation can be mapped onto the corresponding spectrum for equation (1) as

$$W_n = \pm mc^2 \sqrt{1 + \frac{2E_n}{mc^2}}. \tag{14}$$

The exact form of the Schrödinger spectrum depends on the boundary conditions imposed for the particular problem at hand. If the bound-state wave functions are not supposed to vanish in the origin, one may apply the standard set of bounded quasi-polynomial solutions. Though such solutions are not permissible in the 3D quantum mechanics [14], however, they may be useful for certain applications (e.g., [15]). For this case the Hermite functions involved in the expansion of the bi-confluent Heun function become the Hermite polynomials and the Schrödinger spectrum is (exactly) given as [9]

$$E_n = \frac{V_0}{2} \left( \frac{-mV_0}{\hbar^2} \right)^{1/3} \frac{1}{n^{2/3}}, \quad n = 1, 2, 3, \ldots \tag{15}$$

(we recall that $V_0 < 0$).

In the case when one demands the wave function to vanish both in the origin and at the infinity [14], that is if $\psi(0) = 0$ and $\psi(+\infty) = 0$, it is shown in [9] that the Schrödinger spectrum is with high accuracy approximated as

$$E_n \approx \frac{V_0}{2} \left( \frac{-mV_0}{\hbar^2} \right)^{1/3} \left( n - \frac{1}{6} \right)^{-2/3}, \quad n = 1, 2, 3, \ldots, \tag{16}$$

the Maslov correction index being $\approx -1/6$ [16,17].

To discuss the properties of the corresponding energy spectrum for the relativistic wave equation (1), we note that the inverse-square-root potential can be represented in terms of a certain spatial characteristic length $d$ as

$$V(x) = -mc^2 \frac{\lambdabar}{\sqrt{d}} \frac{1}{\sqrt{x}}, \tag{17}$$

where $\lambdabar = \hbar/(mc)$ is de Broglie wavelength of the particle and $d \equiv \left( -mc^2 \lambdabar / V_0 \right)^2$. The two physically feasible scalar potentials for the case of zero vector potential are

$$q_0 \varphi(x) = -mc^2 \frac{2\lambdabar/\sqrt{xd}}{1 \pm \sqrt{1 - 2\lambdabar/\sqrt{xd}}}. \tag{18}$$

With the Schrödinger spectrum (15) for quasi-polynomial bound-state wave functions, from equation (14) we obtain the energy spectrum for equation (1) as

$$W_n = -mc^2\sqrt{1 + \frac{2E_n}{mc^2}} = -mc^2\sqrt{1 - \left(\frac{\lambdabar}{nd}\right)^{2/3}}, \quad (19)$$

It is immediately seen from this equation that the bound states exist only if

$$nd > \lambdabar. \quad (20)$$

It is understood that this relation is a direct reflection of uncertainty principle, which states that a particle cannot be localized in a spatial region smaller than the de Broglie wavelength of the particle. A further observation is that $n \geq n_0$, where $n_0$ is a positive integer defined as

$$n_0 = \lceil \lambdabar/d \rceil, \quad (21)$$

where the upper-corner brackets denote the ceiling function. This restriction shows that the ground-state wave function for equation (1) generally is not single-peaked. Indeed, according to the well appreciated oscillation theorem [18], the $n_0$-th bound state wave function for the Schrödinger equation has $n_0 - 1$ nodes (if the ground state is labeled as the first bound state). Since it is the $n_0$-th bound-state wave function of the Schrödinger equation that corresponds to the ground state wave function for equation (1), we conclude that the number of extremums of the ground-state wave function for equation (1) is conditioned by the characteristic spatial size $d$. We have an *infimum* cut (a cut from below). This is an interesting feature not faced so far in the case of other known relativistic or non-relativistic wave equations (note that in the Schrödinger case the ground-state wave function for monotonic singular potentials is always single-peaked).

Constructing the spectrum for equation (1) for the case of wave functions vanishing both in the origin and at the infinity, by substitution of (16) into (14) and taking into account the physical representation (17) of the inverse-square-root potential, we obtain

$$W_n \approx -mc^2\sqrt{1 + \frac{2E_n}{mc^2}} = -mc^2\sqrt{1 - \left(\frac{\lambdabar}{d(n-1/6)}\right)^{2/3}}. \quad (22)$$

Hence, in this case the infimum cut occurs with

$$n_0 = \left\lceil \frac{\lambdabar}{d} + \frac{1}{6} \right\rceil, \quad n = n_0, n_0+1, n_0+2, n_0+3,.... \quad (23)$$

We conclude this section by noting that there may be other physical problems with different boundary conditions imposed for the origin. It is understood that in these cases the limitation on the limiting ground-state number will change to $n > n_0 = \lceil \lambdabar/d - i_M \rceil$, where $i_M$ is the corresponding Maslov index associated with the particular problem at hand.

## 5. Discussion

Thus, we have transformed the stationary relativistic wave equation for a spinless particle (1) to the one-dimensional stationary Schrödinger equation and reduced the latter equation to the bi-confluent Heun equation by an appropriate transformation of the independent and dependent variables. Physical potentials $\varphi$, $\mathbf{A}$ and energy spectrum $W_n$ for the wave equation under consideration are related to the bi-confluent Heun potential $V$ and corresponding discrete spectrum $E_n$.

We have seen that all the potentials $V(x)$ for one-dimensional stationary Schrödinger equation are also applicable for the considered equation. Notation (3) shows that the variety of physical potentials $\varphi$ and $\mathbf{A}$ for this equation is richer as compared to the Schrödinger case. We note that in the case of absence of the vector potential $\mathbf{A}$ there are always two scalar potentials $\varphi$ for each of the Schrödinger potential $V$.

We have presented the five bi-confluent Heun potentials which are prominent in that they generalize all the classical hypergeometric potentials (i.e., the harmonic oscillator [19], Coulomb [20], Kratzer [21] and Morse [22] potentials) which have been widely applied in establishing and developing the quantum mechanics. Further, we have considered the inverse-square-root bi-confluent Heun potential $V = V_0/\sqrt{r}$ and presented the corresponding form of physically feasible potentials. We have discussed the properties of the bound-states' energy spectrum both for the wave functions vanishing only at infinity and for the wave functions vanishing both at the infinity and at the origin. We note that for the Schrödinger spectrum the spatial and interaction parameters do not impose whatever restrictions on the energy spectrum, including the ground state. In contrast to this, we have seen from equation (14) that the ground state for equation (1) is limited by the spatial characteristic length. This observation directly reflects the Heisenberg uncertainty principle.

The condition for spatial parameter limitation has been shown to be $nd > \lambdabar$. As a consequence, the bound state's possible quantum number is restricted to values higher than those typical of Schrödinger equation, i.e., the ground-state's number starts from $n_0 = \lceil \lambdabar/d \rceil$. Such an infimum cut for physically realizable states is an interesting feature not faced for other known relativistic or non-relativistic wave equations.

**Acknowledgments**

The work was supported by the Armenian State Committee of Science (SCS Grant No. 18RF-139), the Armenian National Science and Education Fund (ANSEF Grant No. PS-4986), and the Russian-Armenian (Slavonic) University at the expense of the Ministry of Education and Science of the Russian Federation